Methane Formation Efficiency on Icy Grains: Role of Adsorption States

Masashi Tsuge[1,2], Germán Molpeceres[3,4], Yuri Aikawa[3], Naoki Watanabe[1]

Abstract

Methane ($CH_4$) is one of the major components of the icy mantle of cosmic dust prevalent in cold, dense regions of interstellar media, playing an important role in the synthesis of complex organic molecules and prebiotic molecules. Solid $CH_4$ is considered to be formed via the successive hydrogenation of C atoms accreting onto dust: $C + 4H \rightarrow CH_4$. However, most astrochemical models assume this reaction on the ice mantles of dust to be barrierless and efficient, without considering the states of adsorption. Recently, we found that C atoms exist in either the physisorbed or chemisorbed state on compact amorphous solid water, which is analogous to an interstellar ice mantle. These distinct adsorption states considerably affect the hydrogenation reactivity of the C atom. Herein, we elucidate the reactivities of physisorbed and chemisorbed C atoms with H atoms via sequential deposition and co-deposition processes. The results indicate that only physisorbed C atoms can produce $CH_4$ on ice. Combining this finding with a previous estimate for the fraction of physisorbed C atoms on ice, we determined the upper limit for the conversion of C atoms into $CH_4$ to be 30%.

Unified Astronomy Thesaurus concepts: Astrochemistry (75); Molecualr clouds (1072); Interstellar molecules (849); Interstellar dust (836); Laboratory astrophysics (2004)

---

[1] Corresponding author tsuge@lowtem.hokudai.ac.jp
[2] Institute of Low Temperature Science, Hokkaido University, Sapporo 060-0819, Japan
[3] Department of Astronomy, Graduate School of Science, The University of Tokyo, Tokyo 113-0033, Japan
[4] Departamento de Astrofísica Molecular, Instituto de Física Fundamental (IFF-CSIC), Madrid 28006, Spain

## 1. Introduction

Methane (CH$_4$), the simplest saturated hydrocarbon, pervades interstellar media. Solid CH$_4$ has been detected towards high- and low-mass young stellar objects (Boogert et al. 1996; Öberg et al. 2008; Rocha et al. 2024), and recently, using the James Webb Space Telescope, scientists identified CH$_4$ in highly extinct molecular clouds (McClure et al. 2023). These observations indicate that the abundance of solid CH$_4$ ranges from 2%–8% as compared to the most abundant solid H$_2$O. In the chemical evolution towards complex organic molecules (COMs) and prebiotic molecules, CH$_4$ is believed to play a key role via photolysis; this hypothesis is supported by observational evidence suggesting the involvement of CH$_4$ in the conversion of an organic inventory containing the compound into COMs in interstellar clouds (Öberg & Bergin 2021). Conversely, CH$_4$ being chemically stable, does not undergo many two-body reactions on the surface of dust grains. Therefore, the CH$_4$ inventory of dust grains is a key unknown in the production of COMs.

In the dense regions of a molecular cloud, with the temperature as low as 10 K, CH$_4$ production is often assumed to occur on icy grain surfaces via the successive hydrogenation of C atoms:

$$C \xrightarrow{H} CH \xrightarrow{H} CH_2 \xrightarrow{H} CH_3 \xrightarrow{H} CH_4. \qquad (1)$$

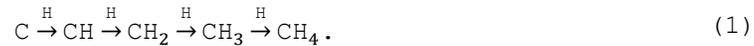

These barrierless and exothermic reactions have been assumed to proceed efficiently on icy grain surfaces in astrochemical models without experimental validation. Recently, laboratory experiments by Qasim et al. (2020a) demonstrated CH$_4$ production via this pathway on a cold substrate. Their co-deposition experiments involved the simultaneous deposition of H atoms, C atoms, and H$_2$O onto a 10 K amorphous carbon substrate, similar to other laboratory experiments on C-atom reactions with potential ice mantle components (Fedoseev et al. 2022; Molpeceres et al. 2021; Potapov et al. 2021). However, these simulation experiments may not fully reflect the chemistry occurring in molecular clouds or protoplanetary disks, where atoms and molecules are slowly transferred from the gas phase onto icy grains. Adsorbates remain on the surfaces of these icy grains for a long time before encountering other species, in contrast to co-deposition experiments where rapid deposition rates facilitate immediate reactions. To extrapolate laboratory experiments to realistic environments, experimentally determining each elementary physicochemical process, such as adsorption, surface diffusion, and reaction, and integrating these findings into astrochemical models is necessary.

In addition to $CH_4$ formation on icy grains, the immediate chemisorption of all C atoms onto water ice has also been proposed on the basis of quantum chemical calculations predicting binding energies (Duflot et al. 2021; Ferrero et al. 2024; Molpeceres, et al. 2021; Shimonishi et al. 2018). When C atoms are chemisorbed, they form C–$OH_2$ or [$COH^-$ + $H_3O^+$] pair or are chemically converted to formaldehyde ($H_2CO$) (Molpeceres, et al. 2021). However, we recently found that not all C atoms are chemisorbed, and a fraction of C atoms remains physisorbed, leading to surface diffusion at low temperatures (Tsuge et al. 2023). The C-atom adsorbates on water ice have been classified into three types:

- **(i)** C atoms that readily react with $H_2O$ to form $H_2CO$,
- **(ii)** C atoms that remain physisorbed for a prolonged period,
- **(iii)** Physisorbed C atoms that transform into the chemisorbed state within ~20 min.

Approximately 30% of C atoms are of type (i), with the remainder belonging to either type (ii) or (iii). The fractions of types (ii) and (iii) are coverage dependent; under molecular-cloud conditions (Tsuge et al. 2023), the type (ii) fraction of C atoms is as high as 40%, declining to ~7% under high coverage (~0.1) conditions. Although clear explanations based on experiment or theory are still lacking, this dependence likely arises from adsorption dynamics, wherein deposited C atoms are unevenly distributed among binding sites, with a limited number of binding sites favoring type (ii) occupation over type (iii) occupation. Another explanation is that C atom distribution is determined among various physisorption sites because of non-negligible barrier to enter chemisorbed states. Between the transformation from a physisorbed state to a chemisorbed state, reactivity may undergo drastic changes because of the formation of covalent C–O bonds. For example, while the C + $H_2$ → $CH_2$ reaction is observed in helium nanodroplets at 0.4 K (Kranokutski et al. 2016), the C–$OH_2$ + $H_2$ → $CH_2$–$OH_2$ reaction has a sizable barrier (27–30 kJ mol$^{-1}$) preventing its occurrence at 10 K (Lamberts et al. 2022). Although experimental evidence on the reactivity of chemisorbed C atoms with H atoms is lacking, theoretical calculations by Ferrero et al. (2024) suggested that the conversion of C–$OH_2$ to $CH_4$ or methanol ($CH_3OH$) depends upon factors such as the binding site and spin multiplicity of the reaction. Determining whether chemisorbed C atoms react with H atoms is critical for accurately modeling the abundance of $CH_4$ in icy mantles.

In this study, we elucidated the reactivities of physisorbed and chemisorbed C atoms and found that only physisorbed C atoms reacted with H atoms to yield $CH_4$. Based on this finding, we determined that a maximum of 30% of C atoms absorbed on ice at low temperatures can be transformed into $CH_4$ under molecular cloud conditions.

## 2. Methods

### 2.1 Apparatus

All experiments were performed using the Reaction Apparatus for Surface Characteristic Analysis at Low-temperature (RASCAL) at the Institute of Low Temperature Science, Hokkaido University. A detailed description of RASCAL has been provided in the literature (Hama et al. 2012; Watanabe et al. 2010). RASCAL consists of an ultrahigh vacuum main chamber (with a base pressure of ~$10^{-8}$ Pa) and differentially pumped chambers for H atom and C atom sources. The aluminum (Al) substrate was positioned at the center of the main chamber and could be cooled to ~6 K using a closed-cycle He refrigerator (RDK-415E, SHI). Temperature regulation of the substrate ranging from 6–300 K was achieved using a silicon-diode temperature sensor (DT-670, Lake Shore), ceramic heaters (MC1020, Sakaguchi E. H VOC), and a temperature controller (Model 335, Lake Shore).

The C-atom source was procured from MBE-Komponenten (Model SUKO-A). This source exclusively emits C[1] in its electronic ground state; hence, no C atom clusters such as $C_3$ are emitted. The flux of the C atom was controlled by adjusting the temperature of the filament, which comprised a thin Ta tube filled with carbon powder. Under the experimental conditions employed in this study, the flux was determined to be (7–8) × $10^{10}$ atoms $cm^{-2}$ $s^{-1}$ according to the method developed by Qasim et al. (2020b). The total amount of deposited C atom was approximately (1.3–1.4) × $10^{14}$ atoms $cm^{-2}$ in all the experiments of this study.

H atoms were generated via the dissociation of $H_2$ molecules in a microwave discharge plasma in a Pyrex tube. Consequently, a gas mixture of H and $H_2$ was generated and transferred to the main chamber via a series of polytetrafluoroethylene tube and an Al tube cooled to 100 K in a closed-cycle He refrigerator. The fraction of dissociation at the substrate surface was ~10%, and the H atom flux was estimated to be (3–5) × $10^{13}$ atoms $cm^{-2}$ $s^{-1}$ according to a previously reported procedure (Hidaka et al. 2007).

## 2.2 Preparation of np-ASW sample

The nonporous amorphous solid water (np-ASW) samples were produced via background vapor deposition of $H_2O$ molecules at a substrate temperature of 110 K. The thickness of np-ASW was approximately 20 monolayers (1 monolayer ≈ 0.3 nm). Subsequently, the np-ASW sample was cooled to 10 K for the deposition experiments. Fresh np-ASW samples were prepared for each experiment to maintain consistency and accuracy.

## 2.3 Infrared measurements

The infrared (IR) spectra of the samples on the Al substrate were measured using reflection-absorption IR spectrometry with a Fourier-transform infrared spectrometer (Spectrum One, Perkin Elmer) equipped with a KBr beam splitter and a Hg–Cd–Te detector. Spectra were collected in a region of 700–4000 $cm^{-1}$, with a correction applied at a resolution of 4 $cm^{-1}$ after averaging 200 or 400 scans.

The surface number density (column density) of species X, denoted as [X] in molecules $cm^{-2}$, was estimated using the following equation:

$$[X] = \frac{\cos\theta \times \ln 10}{2B} \int A(\tilde{v}) d\tilde{v}, \qquad (2)$$

where $\theta$, $B$, and $A(\tilde{v})$ represent the incident angle of the IR beam with respect to the Al substrate (83°), integrated absorption coefficient (in the unit of cm $molecule^{-1}$), and absorbance at a given wavenumber, respectively. For $B$ values, we adopted those previously reported for pure solids for CO and $H_2CO$, and for $CH_4/H_2O$ mixed ice, we used the reported value for $CH_4$. These values are as follows: $1.1 \times 10^{-17}$ for CO (2153 $cm^{-1}$, (Jiang et al. 1975)), $9.6 \times 10^{-18}$ for $H_2CO$ (1720 $cm^{-1}$, (Schutte et al. 1993)), and $4.7 \times 10^{-18}$ for $CH_4$ (1304 $cm^{-1}$, (Boogert et al. 1997)). Note that these $B$ values were obtained using transmission infrared spectrometry, whereas our experiments used reflection-absorption infrared spectrometry. This difference could be a source of uncertainty in estimating the number densities.

## 2.4 Temperature-programmed desorption measurements

We conducted temperature-programmed desorption (TPD) measurements by employing a quadrupole mass spectrometer (Q-MS, M-201QA-TDM, Canon ANELVA Co.). The substrate was warmed from 10 K to 180 K with a ramping rate of 5 K $min^{-1}$. To determine the surface number density of $H_2CO$, we converted the integrated intensity of the TPD spectrum (*m/z* 29) assuming that 30% of the deposited C atoms, which amounted to $4 \times 10^{13}$ atoms $cm^{-2}$, were converted to $H_2CO$, according to the observations of Molpeceres et al. (2021). The surface number density of CH$_4$ was determined using the integrated intensity of *m/z* 15 channel, conversion factor used for H$_2$CO, ionization efficiency for CH$_4$ and H$_2$CO (Kim et al.), and fragmentation patterns of CH$_4$ and H$_2$CO (Linstrom & Mallard).

## 3. Results and Discussion
### 3.1 Product identification via infrared spectroscopy

Several experiments were performed, as listed in Table 1. In all experiments, np-ASW samples served as analogs of ice mantles, and C and H atoms were deposited at 10 K. We identified the reaction products using IR spectroscopy and determined the formation mechanisms of these products.

The IR spectra measured after the co-deposition of the C and H atoms (Expt. 1.1) and, C atoms and H$_2$ (Expt. 1.2) are shown in the top panel of Fig. 1. In both experiments, the $\nu_4$ mode of CH$_4$ was clearly observed around 1304 cm$^{-1}$, consistent with the findings of Qasim et al. (2020a) who identified CH$_4$ in either co-deposition of C + H + H$_2$O or C + H$_2$ + H$_2$O. The formation of CH$_4$ in Expt. 1.1 can be attributed to the successive hydrogenation of the C atoms, whereas in Expt. 1.2, it may have been initiated with a barrierless C + H$_2$ reaction (Lamberts et al. 2022), i.e., $^3$C + H$_2$ → $^3$CH$_2$ followed by further reactions with H$_2$ to become CH$_4$. Although the occurrence of the C + H$_2$ reaction in helium nanodroplets has been reported (Krasnokutski et al. 2016); Lamberts et al. (2022) suggested, on the basis of quantum chemical calculations, that this reaction is suppressed once C atoms are chemisorbed onto water ice. The number density of CH$_4$ was determined to be (1.6 ± 0.1) × 10$^{14}$ and (0.3 ± 0.1) × 10$^{14}$ molecules cm$^{-2}$ in Expt. 1.1 and 1.2, respectively. Note that the number densities determined from the IR measurements may have relatively large uncertainties because we used the integrated IR absorption coefficient determined via transmission spectrometry, whereas we employed reflection-absorption spectrometry. The $\nu_2$ and $\nu_3$ modes of H$_2$CO were observed around 1720 and 1500 cm$^{-1}$, respectively. The IR feature observed at 2140 cm$^{-1}$ originates from carbon monoxide (CO), which is a dominant contaminant from the C-atom source. As the hydrogenation of CO efficiently produces H$_2$CO (Tsuge & Watanabe 2023; Watanabe & Kouchi 2002, 2008), an enhancement of the H$_2$CO intensity in Expt. 1.1 compared with that in Expt. 1.2 can be attributed to the hydrogenation of CO. The amount of CO after the deposition was smaller in Expt. 1.1 than in 1.2. The formation of H$_2$CO in Expt. 1.2 was presumably due to the reaction of the C atom with H$_2$O molecules (Duflot et al. 2021; Molpeceres et al. 2021; Potapov et al. 2021).

We identified three types of C atom adsorbates on ASW (Tsuge et al. 2023) as described in the introduction. To evaluate the contribution of type (iii) C atoms, Expt. 2.1 and 3.1 were conducted. In Expt. 2.1, H-atom deposition immediately followed C-atom deposition, whereas in Expt. 3.1, an ice sample deposited with C atoms was maintained at 10 K for 60 min before H-atom deposition. As shown in the second and third panels from the top of Fig. 1, no significant differences were observed between Expt. 2.1 and 3.1. We first explain the results of Expt. 3.1 in the following. After C-atom deposition, the $\nu_2$ and $\nu_3$ modes of $H_2CO$ were observed (Fig. 1e). In the difference spectrum showing the results of subsequent H-atom deposition (Fig. 1f), a reduction in CO and an increase in $H_2CO$ were observed, indicating that the $H_2CO$ produced during H-atom deposition likely resulted from the successive hydrogenation of CO. The decrease in the number density of CO closely matched the increase of $H_2CO$: 5–6 × $10^{13}$ molecules cm$^{-2}$. Importantly, the number density of $CH_4$ fell below the detection limit (presumably, <1 × $10^{13}$ molecules cm$^{-2}$). Considering that a 60 min time interval between C-atom deposition and H-atom deposition is sufficiently longer than the 20 min timescale for the transformation from the physisorbed state to the chemisorbed state, type (iii) C atoms were converted to the chemisorbed state at the onset of H-atom deposition. Therefore, this finding suggests that chemisorbed C atoms cannot undergo hydrogenation to produce $CH_4$; i.e., the successive hydrogenation reaction, C–$OH_2$ + 4H → $CH_4$ + $H_2O$, either does not occur or occurs with extremely low yield. No significant differences were observed between Expt. 2.1 and 3.1, possibly because the conversion from physisorption to chemisorption occurred during C-atom deposition (see, for example, Fig. S4 of Tsuge et al. (2023)); thus, a substantial fraction of physisorbed C atoms were already transformed into chemisorbed C atoms even upon the termination of C-atom deposition. In Expt. 2.2 and 3.2, no spectral changes were observed after $H_2$ deposition.

There is a mismatch between experiments and theory on the methane formation from chemisorbed carbon. In the following, we discuss the mismatch based on the competition between hydrogenation and reaction with water molecules. To form $CH_2$ from chemisorbed C atoms, two sequential hydrogenation reactions are required (Ferrero et al. 2024). The first hydrogenation reaction C-$OH_2$ + H → HC-$OH_2$ reaction is highly exothermic and the heat of reaction enhances a further reaction with water molecules (proton relay) to form $CH_2OH$. Consequently, the $CH_2OH$ formation would be induced before the second hydrogenation reaction, meaning that the hydrogenation of chemisorbed carbon to $CH_4$ is not plausible under our experimental condition as well as in the molecular cloud conditions, where the timescale of hydrogenation reactions is determined by the adsorption rate of H atoms. Finally, co-deposition experiments, like the ones of Qasim et al. (2020) or our own, can form methane because the encounter rate of C and H atoms is greatly enhanced in these experiments.

To discriminate the $H_2CO$ formation from mechanisms other than the CO hydrogenation, Expt. 4 was performed. In this experiment, after the deposition of C atom at 10 K, the sample underwent annealing at 60 K for a few minutes, followed by H-atom deposition after cooling the sample to 10 K. The difference spectrum post-annealing (Fig. 1h) clearly exhibited a decrease in the number density of CO. Conversely, in the difference spectrum representing the result of H-atom deposition (Fig. 1i), no spectral changes were observed, indicating that reactive CO molecules were eliminated from the sample upon annealing at 60 K. Although a significant fraction (50%–60%) of CO remained in the sample even after annealing, these were considered to have migrated into ASW (He et al. 2018; Karssemeijer et al. 2014; Lauck et al. 2015; Mispelaer et al. 2013), where the reaction with H atoms is inefficient at 10 K (Tsuge et al. 2020). Hence, from this experiment, we conclude that the production of $H_2CO$ during H-atom deposition in Expt. 2.1 and 3.1 primarily arose from the successive hydrogenation of CO.

The formation of $CH_3OH$ was not observed in all measurements. Lamberts et al. (2022) tentatively identified $CH_3OH$ formation in $H_2O$/C/H/$H_2$ co-deposition experiments and they attributed it to the successive hydrogenation of $H_2CO$. The non-detection of $CH_3OH$ in our experiments would result from the low flux of H atom that cannot induce the conversion of CO and $H_2CO$ to $CH_3OH$; the fluence employed in this work was (5–9) × $10^{16}$ atoms $cm^{-2}$, which is much smaller than the fluence (>1 × $10^{18}$ atoms $cm^{-2}$) adopted in other CO hydrogenation experiments (e.g., Watanabe & Kouchi 2002). Recently, theoretical calculations by Ferrero et al. (2024) indicated that C–$OH_2$ (C atoms chemisorbed on $H_2O$) reacts with H atoms to yield $CH_3OH$. In this study, Expt. 4 is the most appropriate for investigating this process owing to the suppression of $CH_3OH$ formation via the successive hydrogenation of CO. As depicted in Fig. 1i, the anticipated IR signatures of $CH_3OH$ in the 1035–1015 $cm^{-1}$ range were not observed. Considering that the integrated absorption coefficients of $CH_3OH$ and $H_2CO$, $CH_3OH$ should have been observed after H-atom irradiation in Expt. 2.1, 3.1, and 4, even with the assumption that only 10% of C–$OH_2$ was converted to $CH_3OH$. This finding diverges from theoretical predictions by Ferrero et al. (2024). According to Molpeceres et al. (2021) and Ferrero et al. (2024), the formation of $CH_3OH$ involves proton relays with the water molecule in highly diluted conditions (also observed in Molpeceres et al. (2024) for C-$NH_3$). However, when the concentration of C-$OH_2$ on the surface is larger, the proton relays for C-$OH_2$ → HCOH reaction might be partially inhibited or the product for the reaction is not suitable for the formation of $CH_3OH$. We anticipate that this reduced efficiency, summed to the insufficient H atom flux for $CH_3OH$ formation discussed above, makes the $CH_3OH$ formed in the reaction mixture hardly detectable in the experiments.

### 3.2 Quantification of methane yields via temperature-programmed desorption

The accurate determination of $CH_4$ and $H_2CO$ yields poses challenges owing to uncertainties in calculating number densities from IR spectra. However, in TPD experiments, detection via Q-MS offers higher sensitivity compared with that via IR spectroscopy. Consequently, the relative yields of $CH_4$ and $H_2CO$ can be determined more accurately. A key challenge with TPD experiments lies in establishing the absolute number density of the adsorbates. Here, we assumed that 30% of the deposited C atoms underwent rapid conversion to $H_2CO$, as proposed by Molpeceres et al. (2021). During the TPD measurements, we determined the $CH_4$ and $H_2CO$ yields by analyzing the signals from $m/z$ 15 and 29 channels, respectively. This analysis involved considerations of relative ionization efficiency and fragmentation patterns. The determined yields are listed in Table 1.

Figure 2 illustrates the TPD spectra of $m/z$ 29 and $m/z$ 15 channels recorded in Expt. 1–5. In Fig. 2a, the TPD signal of $H_2CO$ ($m/z$ 29) in the temperature range of 95–115 K was observed across all the experiments. The signals observed in Expt. 1.2, 2.2, 3.2, 4, and 5 indicate the yield of $H_2CO$ directly from the C + $H_2O$ reaction. For subsequent analysis, these signal intensities were presumed to originate from $H_2CO$ formed from 30% of the deposited C atoms, corresponding to a number density of ~0.4 × $10^{14}$ molecules $cm^{-2}$. In the H-atom deposition experiments (Expt. 1.1, 2.1, and 3.1), $H_2CO$ could have been produced via the reaction of C + $H_2O$ and/or via the successive hydrogenation of CO. As shown in Fig. 2a and listed in Table 1, the signal intensities in Expt. 2.1 and 3.1 were marginally enhanced compared with those in Expt. 1.1. This enhancement can be rationalized by considering that the fraction of the type (i) C atoms in Expt. 1.1 was lower than that in Expt. 2.1 and 3.1. In other words, in Expt. 1.1, C atoms were rapidly consumed via hydrogenation before participating in the C + $H_2O$ reaction, that leads to the formation of $H_2CO$. The $H_2CO$ number density in Expt. 1.1 was determined to be $(1.3 \pm 0.2) \times 10^{14}$ molecules $cm^{-2}$, whereas that in Expt. 2.1 and 3.1 was $(1.7 \pm 0.1) \times 10^{14}$ molecules $cm^{-2}$ (Table 1). The discrepancy of $\sim 0.4 \times 10^{14}$ molecules $cm^{-2}$ corresponds to approximately 30% of the deposited C atoms, aligning with our assumption that 30% of C atoms directly contribute to the formation of $H_2CO$. This result suggests that the presence of $H_2CO$ in Expt. 1.1 was solely from the successive hydrogenation of CO contaminants.

As depicted in Fig. 2b, Expt. 1.1 exhibited a prominent signal of $CH_4$ ($m/z$ 15). The TPD signal of $CH_4$ exhibited an increase beginning at 35 K, reaching its maximum at approximately 50 K, and then decreasing to the baseline level at 65 K. The number density of $CH_4$ was calculated to be $(1.5 \pm 0.2) \times 10^{14}$ molecules $cm^{-2}$. This value corresponds with the fluence of the C atom, estimated at $(1.3–1.4) \times 10^{14}$ atoms $cm^{-2}$, within the statistical margin of error. This alignment indicates that the majority of the deposited C atoms underwent conversion to $CH_4$ during the co-deposition of H and C atoms. In Expt. 1.2, consistent with the IR measurements, the TPD signal for $CH_4$ was weaker. In other experiments, except for Expt. 4, where $CH_4$ was undetectable, a minor quantity of $CH_4$, ranging between $(3–7) \times 10^{12}$ molecules $cm^{-2}$, was observed. This small amount of $CH_4$ most likely originated from the reaction of type (ii) C atoms with H atoms or $H_2$. The fraction of type (ii) C atoms under these conditions is less than 7% of the total amount of deposited C atoms (Tsuge et al. 2023). To establish an upper limit, we followed the estimates by Molpeceres et al. (2021), where approximately 30% of the deposited C atoms were readily converted to $H_2CO$. The C atom coverage-dependent branching ratio between types (ii) and (iii) C atoms has been experimentally determined (Tsuge et al. 2023) to be 0.1 and 0.9 for types (ii) and (iii), respectively, under high coverage conditions. Thus, the fraction of type (ii) C atoms is derived to be 7% (70% × 0.1). As discussed in the introduction, the fraction of type (ii) C atoms decreases with increasing coverage and can reach up to 40% under fluences relevant to molecular cloud conditions, as elaborated in the subsequent section. In terms of number density, less than 7% of deposited C atoms roughly corresponds to $<1 \times 10^{13}$ molecules $cm^{-2}$. This alignment with the estimated number density of $CH_4$, $(3–7) \times 10^{12}$ molecules $cm^{-2}$ (Table 1), lends support to the scenario where $CH_4$ observed in Expt. 2, 3, and 5 originated from the reaction of type (ii) C atoms with H atom or $H_2$. Although H atom or $H_2$ were not intentionally introduced in Expt. 5, the residual $H_2$ molecules that accumulated on the substrate during the experiment may have facilitated $CH_4$ formation. In Expt.

4, CH$_4$ signals were not observed, possibly because any CH$_4$ formed during C-atom deposition was desorbed during the annealing process and physisorbed C atoms were consumed in diffusive recombination reactions (Tsuge et al. 2023).

Because CO is a dominant contaminant from the C-atom source and we adopted a relatively high C-atom coverage in the experiments, there is a possibility that C atom react with CO to produce CCO, which would be converted to ketene (CH$_2$CO) upon hydrogen irradiation (Fedoseev et al. 2022). To exclude this possibility, we performed experiments similar to Expt. 3.1. In Fig. 3, TPD spectra of *m/z* 15, 29, and 42 channels for the temperature range 20–180 K are presented. For the *m/z* 42 channel characteristic for CH$_2$CO, practically no increase in signal intensity was observed near the desorption temperature of CH$_2$CO (88–107 K) and upon ice sublimation (>150 K). Other potential sinks of C atoms initiated with the C + CO reaction are also discussed in Appendix and we conclude that these chemical routes are negligible in our experiments.

### 3.3 Astrophysical implication

The experimental results presented herein strongly suggest that, on the ASW surface, only physisorbed C atoms will lead to the formation of CH$_4$ via reactions initiated by H or H$_2$. Once the C atoms undergo chemisorption, they cannot form CH$_4$ upon irradiation with H or H$_2$. Incorporating this finding into our previously reported framework (Tsuge et al. 2023), we have developed an updated scenario in Fig. 4 illustrating the accretion of C atoms onto ASW at low temperatures (~10 K). The fractions of the total C atoms are given as percentages as follows.

Upon deposition, all the C atoms initially undergo physisorption (with a binding energy (BE) < 5000 K), with approximately 30% of the atoms readily reacting with H$_2$O to produce H$_2$CO. The fraction (30%) aligns with the theoretical prediction by Molpeceres et al. (2021), while our experimental results corroborate this estimate as reasonable. Approximately 40% of the C atoms form C–OH$_2$, indicative of chemisorption with a BE > 10000 K, within a timescale of $10^3$–$10^4$ s. Once chemisorbed, this fraction remains inert toward reactions with H or H$_2$. Our experiments indicate a lack of CH$_4$ formation, yet we cannot exclude the production of other products present at concentrations below the detection limit. The transformation from the physisorbed to the chemisorbed states is hypothesized to proceed via quantum mechanical tunneling, with both the fraction (40%) and timescale ($10^3$–$10^4$ s) being experimentally determined (Tsuge et al. 2023). Consequently, the remaining 30% of the deposited C atoms persisted in the physisorbed state, representing the upper limit for CH$_4$ production via the successive hydrogenation of C atoms. While the true chemical nature of the physisorbed C atoms with BE < 5000 K is not yet complete clear, they should correspond to intermediate situations between what was reported in Das et al. (2018) for pure, very weakly bound C atoms and the chemisorbed carbon atoms reported in the astrochemical literature (Duflot et al. 2021; Ferrero et al. 2024; Molpeceres, et al. 2021; Shimonishi et al. 2018).

Astronomical observations indicate that the estimated solid CH$_4$ abundance with respect to solid H$_2$O is 2%–8% (McClure et al. 2023; Öberg et al. 2008), whereas models predict 10%–24% (Garrod & Herbst 2006). Under the assumption that C atoms are efficiently converted to CH$_4$, this low observational value has been explained as most of C atoms are already locked up in CO, in the gas phase, at the time for CH$_4$ formation (Hiraoka et al. 1998; Öberg et al. 2008). However, the C atom budget can be reconciled with the observed CH$_4$ yield by considering the fraction of C atoms converted to CH$_4$. Our study supports the theoretical predictions, suggesting that approximately 30% of the C atoms readily converted to H$_2$CO, with an upper limit of 30% of C atoms for CH$_4$ production. Consequently, updating astrochemical models can lead to more accurate predictions for CH$_4$ and, eventually, for the organic inventory. Notably, a model calculation by Garrod and Herbst indicates a CH$_4$ abundance of 10%–24% in the total ice during the collapse phase (Garrod & Herbst 2006). By incorporating the upper limit (30%) determined in our study, the predicted CH$_4$ abundance can be reduced to 3%–8%, thus aligning models more closely with observations for this specific molecule.

Non-hydrogenated or partially hydrogenated species, such as C, CH, CH$_2$, and CH$_3$, undergo diffusive reactions as the surface temperature increases during star formation. Among these species, the activation energy for C atom diffusion in ASW has been previously reported (Tsuge et al. 2023). Considering the activation energy and timescale relevant to the lifecycle of icy grains, we previously suggested that at approximately 22 K, the diffusion of C atoms becomes activated, leading to C-atom insertion and addition reactions that induce skeletal evolution, such as C–C bond formation, of molecules toward COMs or carbon-chain molecules (Tsuge et al. 2023). We note that diffusive reactions of other light species, such as O atom, N atom, and CO, will be activated at temperatures below 22 K; thus, these species might react with physisorbed C atoms. For example, the product of C-atom reaction with CO, CCO radical, and its hydrogenated product CH$_2$CO have been suggested to be an important in the formation of COMs on the icy grains in molecular clouds (Fedoseev et al. 2022). Among C-bearing species other than the C atom, the activation energies for $CH_2$ and $CH_3$ diffusion on ASW are considered important for improving the predictions made by astrochemical models (Furuya et al. 2022), and we are currently developing a method to determine these energies.

The fate of the chemisorbed C atoms, constituting 40% of the deposited C atoms, remains unclear. A recent theoretical calculation suggested that the C–$OH_2$ moiety undergoes reactions with H atoms to produce $CH_2$ or $CH_3OH$ (Ferrero et al. 2024), with the former further reacting with H atoms to produce $CH_4$. Although our experimental findings, i.e., non-reaction of chemisorbed C atoms with hydrogen, might seem contradictory to these calculations, the possible reasons for the mismatch between experiment and theory have been discussed in Section 3.1.

The theoretical exploration of the reactions of the C–$OH_2$ moiety with other radicals, including atoms, such as N($^4$S), O($^3$P), OH, $CH_3$, and $NH_2$, has also been investigated (Ferrero et al. 2024). From an experimental perspective, a method must be developed for the selective detection of chemisorbed C atoms. In this regard, the combination of photostimulated desorption (PSD) and resonance-enhanced multiphoton ionization (REMPI), known as the PSD-REMPI method (Tsuge & Watanabe 2021, 2023), has the potential to tune the photon wavelength for the PSD process to desorb C atoms upon the selective dissociation of the C–$OH_2$ bond.

## 4. Acknowledgments

This study was partially supported by JSPS KAKENHI (Grant No. JP24K00686, JP23H03982, JP22H00159, JP21H01139, JP18K03717, JP20H05847, JP22F22013, and JP17H06087). We acknowledge support from the JSPS International Fellowship Program (Grant No. P22013). G.M acknowledges support from the grant RYC2022-035442-I funded by MCIN/AEI/10.13039/501100011033 and ESF+.

## 5. Appendix

An anonymous reviewer of this paper has suggested that the C-atom reactions with CO, which is a dominant contaminant from the C-atom source, could be a sink of deposited C-atoms. After careful consideration based on our preliminary experiments and calculations performed on response to the suggestion, we have excluded such possibilities. In this appendix, we briefly describe the results.

Because the reactivity of C atoms with CO molecule has been reported (Fedoseev et al. 2022), we firstly discuss a possibility of C-atom consumption by the reaction with CO. The number density of CO molecules after C-atom deposition is approximately $2.5 \times 10^{14}$ atoms cm$^{-2}$, indicating that about 25% of adsorption sites on np-ASW are occupied by CO when assuming CO uniformly covers np-ASW. Consequently, a large fraction of deposited C atoms can interact with CO or is deposited in the vicinity of CO molecule. It has been reported that the C atom reaction with CO leading to CCO is barrierless in the gas phase (Papakondylis & Mavridis 2019), and on CO ice (Ferrero et al. 2023). If this is the case, a significant amount of CCO should be produced during C-atom deposition or CCO should further converted to $C_3O_2$ upon reaction with another CO. CCO should be observed, in IR spectra, near 1988 cm$^{-1}$ (on solid CO, unpublished result) and $C_3O_2$ near 2233 cm$^{-1}$ (in $H_2O/C_3O_2$ = 10/1 ice, Gerakines & Moore (2001)). These IR features are absent in the IR spectra shown in Figure 1. Quantum chemical calculations show that the relation of IR band intensities among CO (75 km mol$^{-1}$), $^3$CCO (141 km mol$^{-1}$), and $C_3O_2$ (2940 km mol$^{-1}$) is clearly in favor of CCO and $C_3O_2$ based on our own calculations at the B3LYP/cc-pVTZ level of theory (Becke 1993; Kendall et al. 1992). The absence of the prominent peaks attributable to CCO and $C_3O_2$ in our experiments indicates that the consumption of C atoms in the reaction with CO should be negligible. We note that barrierless reaction does not guarantee 100% yield upon encounter of reactants. For example, CO + C reaction could be hindered on an ice surface when C atom of adsorbed CO is not exposed. In addition, once C atom is chemisorbed, it will not react with CO (Ferrero et al. 2024). Because of a high surface coverage of C atoms employed in experiments, the formation of $C_2$ via, for example, the Eley–Rideal mechanism is possible. Indeed, we reported the formation of $C_2$ at 10 K (Tsuge et al. 2023), but the fraction is not clear.

Additionally, the CCO radical can be hydrogenated to form ketene, CCO + 2H → $CH_2CO$ (Fedoseev et al. 2022). The reviewer of this manuscript suggested two alternative pathways to form $CH_2CO$. One is the reaction of $CH_2$, an intermediate of the successive hydrogenation of C atom, with CO, and the other is the reaction of CCO with $H_2$ (either from residual gas or intentional deposition). The former path was suggested based on the work by Kranokutski et al. (2017), where the authors suggested, based on their computations, that $CH_2$ + CO → $CH_2CO$ is barrierless. We carried out exploratory calculations in the gas phase, at the BHLYP-D3BJ/ma-def2-TZVP (Becke 1993; Zheng et al. 2010) level of theory and using the ORCA code (Neese 2011; Neese et al.

2020). The reaction of the ground state triplet methylene $^3CH_2$ with CO has a barrier of approximately 10 kJ mol$^{-1}$ whereas that of singlet methylene $^1CH_2$, which is an excited state, is apparently barrierless; based on our potential energy scans; our results are consistent with the literature (King et al. 1998; Ogihara et al. 2010). Because hydrogenation of C atoms on ASW is considered to produce only $^3CH_2$ (Ferrero et al. 2024) and a barrier of 10 kJ mol$^{-1}$ (ca. 1200 K) is high enough to inhibit the reaction, the ketene formation via the $CH_2$ + CO will not be relevant at 10 K.

The CCO + $H_2$ → CHCO + H pathway is further considered here. Our preliminary computations in the gas phase at the BHLYP-D3BJ/ma-def2-TZVP level do show unambiguous endothermicity (~23 kJ mol$^{-1}$); thus, this reaction is not feasible at low temperatures. We also sought for a transition state for the CCO + $H_2$ → $CH_2CO$ concerted reaction, not finding any first order saddle points in the potential energy surface, and making us consider that the reaction is not elementary. We note that the reaction in the ground state of the reactants to $^1CH_2CO$ is spin-forbidden, because the ground state of CCO is $^3\Sigma$. These results are in accordance with the gas-phase study on CCO reactions with H and $H_2$ (Horie et al. 1983). At 298 K, the rate constant of barrierless reaction CCO + H → product(s) is reported to be 4 × 10$^{-11}$ cm$^3$ s$^{-1}$ whereas that of CCO + $H_2$ → product(s) is 7 × 10$^{-13}$ cm$^3$ s$^{-1}$; the difference of about two orders of magnitude strongly indicates that the latter reaction has a barrier.

A final comment pertained to Expt. 4. It was argued by the reviewer that, after heating the sample, the remaining feature near 2150 cm$^{-1}$ could be solely due to $CH_2CO$, and the number density of $CH_2CO$ should be as high as 1.5 × 10$^{13}$ molecules cm$^{-2}$, due to reactions of CO + C + $H_2$. We note here that even in Fedoseev et al. (2022), they showed that a large amount of C$^{18}$O remained in the sample after warming-up to 55 K (see Figure 1d of Fedoseev et al. (2022)). Considering that the number density of $CH_2CO$ assumed by the reviewer, 1.5 × 10$^{13}$ molecules cm$^{-2}$, is about twice as much as the number density of $CH_4$ in sequential deposition experiments, we should be able to detect $CH_2CO$ by TPD measurements. Therefore, we performed additional experiments, similar to Expt. 3.1. We briefly remind that in Expt.3.1 we first deposit C atoms and wait for 60 min before H irradiation. In this H-atom irradiation experiment, $CH_2CO$ formation should be efficient if CCO exists significantly. The TPD spectra for *m/z* 15, 29, and 42 are shown in Figure 3. According to the NIST Chemistry WebBook, *m/z* 42 is the most intense channel in the electron impact ionization of $CH_2CO$ molecule (Linstrom & Mallard). It should be noted that the flux of H-atoms was much higher than in the experiments summarized in Table 1, and therefore, hydrogenation of CO molecules proceeded more to produce higher abundance of $H_2CO$. As shown in Figure 3, the signal characteristic of $CH_2CO$, i.e., *m/z* 42 at temperatures 88–107 K, (Fedoseev et al. 2022) is absent in our experiments. We did not identify $CH_2CO$ via the *m/z* 14 channel either. We note that the *m/z* 42 signal also remained very weak during the desorption of water ice: i.e., at temperatures above 150 K.

Though it has been known that detection of $CH_2CO$ by IR spectroscopy is challenging due to overlapping of most intense band (the $\nu_2$ mode, integrated absorption coefficient $B$ = 1.2 × $10^{-16}$ cm molecule$^{-1}$ (Berg & Ewing 1991)) with CO, we discuss the detectability here. The $B$ values for $\nu_3$ and $\nu_4$ modes expected near 1374 and 971 cm$^{-1}$ are 2.7 × $10^{-18}$ and 4.6 × $10^{-18}$ cm molecule$^{-1}$ (Berg & Ewing 1991), which are comparative to the $B$ value of the $\nu_4$ mode of $CH_4$. In Expt. 1.2, $CH_4$ with the number density of 3 × $10^{13}$ cm$^{-2}$ was clearly observed and, on the other hand, we did not observe any features near 1374 and 971 cm$^{-1}$. Indicating that the $CH_2CO$ yield was much smaller than that of $CH_4$.

In summary, based on all the available experimental evidence, we conclude that CCO, and further products, like $C_2O_3$ or $CH_2CO$, formation via C-atom reactions with CO, is negligible in the current experimental conditions.

**Table 1.** Experiments performed in this study, including measurements of CH$_4$ and H$_2$CO yields determined via IR and TPD analysis.

|   |   | Number density (IR) ($\times 10^{14}$ molecules cm$^{-2}$) | | Number density (TPD)[a] ($\times 10^{14}$ molecules cm$^{-2}$) | |
|---|---|---|---|---|---|
|   | Procedures[b,c] | CH$_4$ | H$_2$CO[d] | CH$_4$ | H$_2$CO[d] |
| 1.1 | C + H co-deposition | 1.6 ± 0.1 | 1.2 ± 0.3 | 1.5 ± 0.1 | 1.3 ± 0.2 |
| 1.2 | C + H$_2$ co-deposition | 0.3 ± 0.1 | 0.4 ± 0.2 | 0.40 ± 0.02 | 0.30 ± 0.05 |
| 2.1 | C atom followed by H atom | ND | 1.9 ± 0.2 | 0.07 ± 0.01 | 1.71 ± 0.05 |
| 2.2 | C atom followed by H$_2$ | ND | 0.5 ± 0.1 | 0.04 ± 0.01 | 0.47 ± 0.01 |
| 3.1 | C atom, 60 min at 10 K, H atom | ND | 1.6 ± 0.3 | 0.07 ± 0.01 | 1.66 ± 0.18 |
| 3.2 | C atom, 60 min at 10 K, H$_2$ | ND | 0.5 ± 0.1 | 0.05 ± 0.02 | 0.42 ± 0.06 |
| 4 | C atom, annealing at 60 K, H atom | ND | 0.6 ± 0.1 | 0.01 ± 0.01 | 0.34 ± 0.03 |
| 5 | C atom | ND | 0.5 ± 0.2 | 0.03 ± 0.01 | 0.38 ± 0.04 |

[a]Based on Molpeceres et al. (2021), 30% of the deposited C atoms were assumed to be converted to H$_2$CO in the absence of H atoms.
[b]Approximate flux and fluence of C atom, H atom, and H$_2$ beams are as follows: C atom, flux = (7–8) × 10$^{10}$ atoms cm$^{-2}$ s$^{-1}$, fluence (30 min) = (1.3–1.4) × 10$^{14}$ atoms cm$^{-2}$; H atom, flux = **(3–5) × 10$^{13}$** atoms cm$^{-2}$ s$^{-1}$, fluence (30 min) = (5–9) × 10$^{16}$ atoms cm$^{-2}$; H$_2$, flux = ~2.5 × 10$^{14}$ molecules cm$^{-2}$ s$^{-1}$, fluence (30 min) = ~5 × 10$^{17}$ molecules cm$^{-2}$.
[c]C atom, H atom, and H$_2$ depositions were performed at a substrate temperature of 10 K for 30 min.
[d]The number densities of H$_2$CO determined from experiments are listed. The H$_2$CO molecules can be formed from C + H$_2$O reaction and hydrogenation of CO contaminants; see text for details.

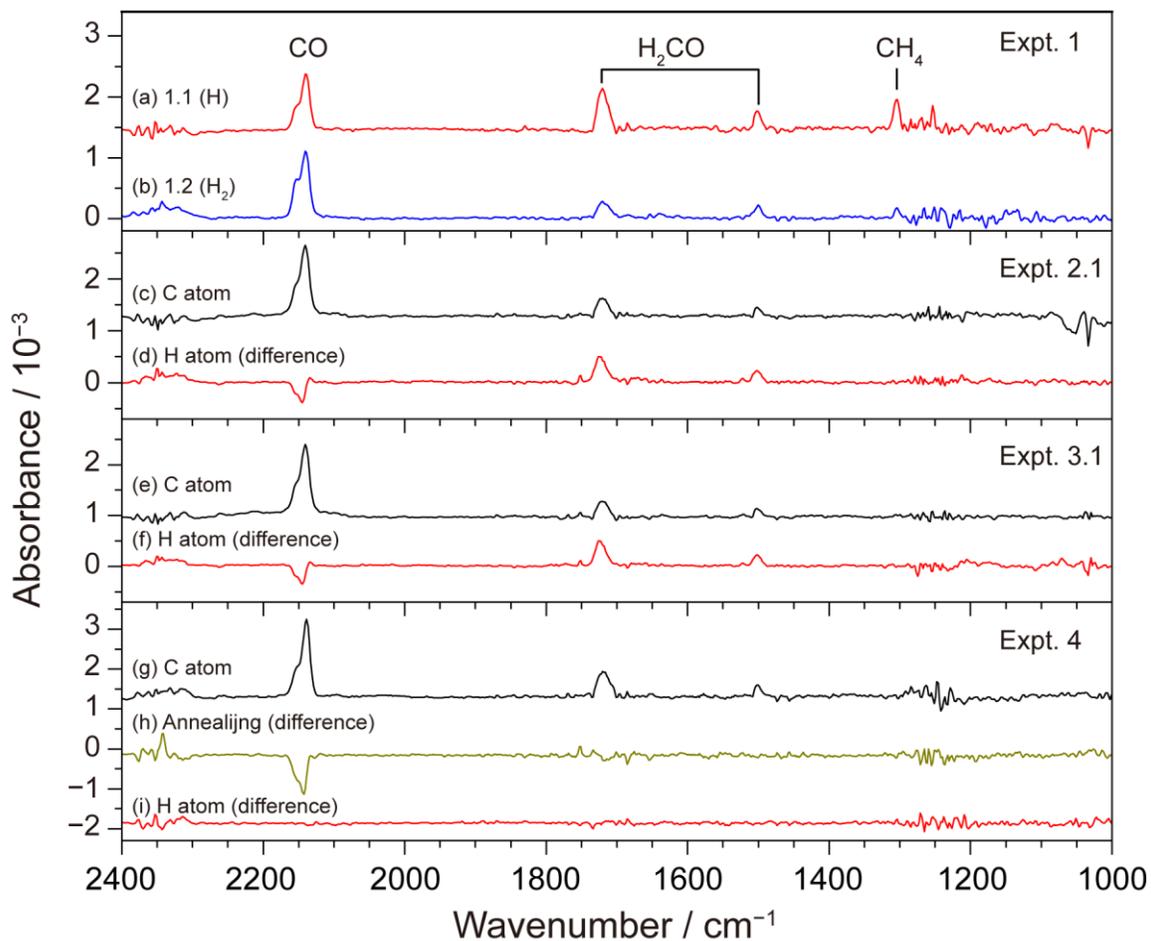

**Figure 1.** Infrared spectra obtained from experiments (from top to bottom) 1.1 and 1.2, 2.1, 3.1, and 4. In the bottom panel, IR features attributed to $CH_4$, $H_2CO$, and CO are marked. Spectra (d), (f), (h), and (i) represent difference spectra illustrating the result of the process indicated by labels; features pointing upward indicate production and those pointing downward indicate destruction.

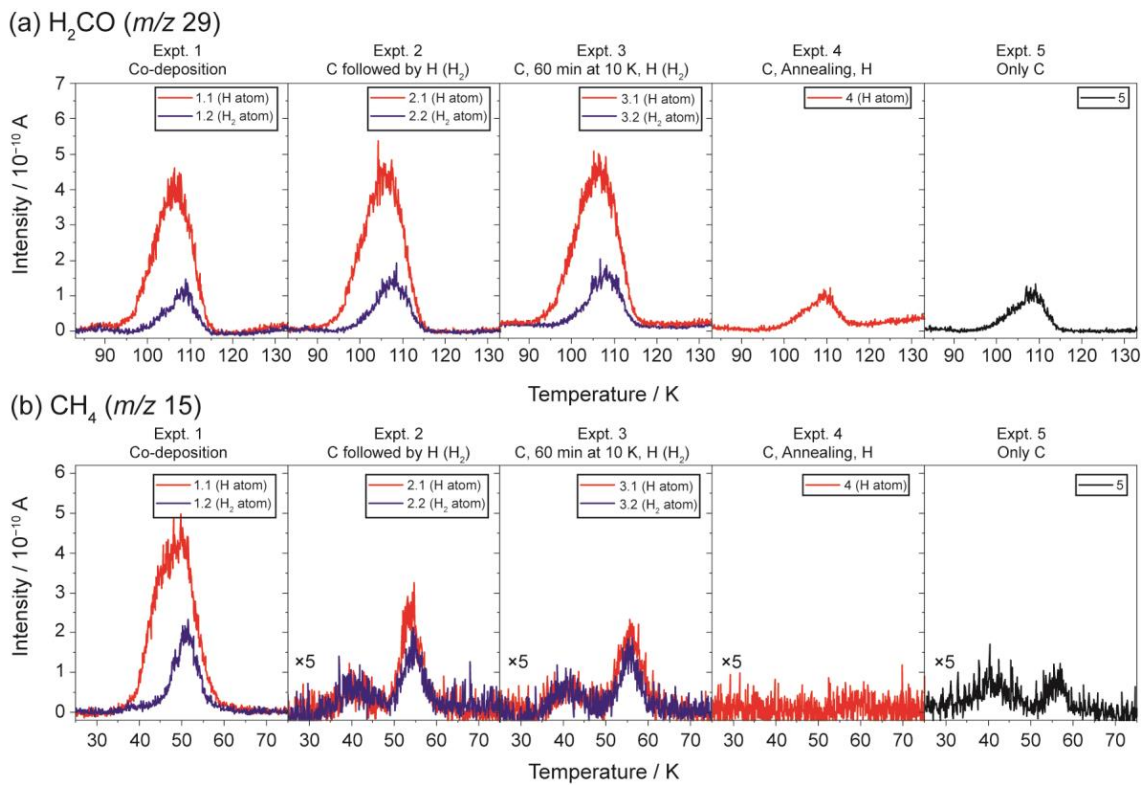

**Figure 2.** Temperature-programmed desorption spectra of (a) H$_2$CO (*m/z* 29) and (b) CH$_4$ (*m/z* 15). The spectra obtained from H atom deposition experiments are depicted in red, those involving H$_2$ deposition are depicted in blue, and only C atom deposition is depicted in black. The TPD traces were baseline-corrected, ensuring a zero background level. In panel (b), all traces except for Expt. 1 were magnified by a factor of 5 for clarity. The *m/z* 15 signal in the range of 35–45 K is due to an interference from a strong signal at *m/z* 16 originating from O$_2$.

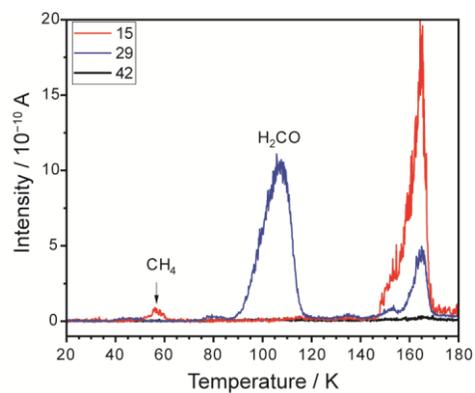

**Figure 3.** Temperature-programmed desorption spectra of *m/z* 15 (red), 29 (blue), and 42 (black) channels obtained from an experiment similar to Expt.3.1 but with a higher H dose. The TPD traces were baseline-corrected, ensuring a zero background level. Peaks due to desorption of $CH_4$ and $H_2CO$ are marked. The TPD trace for the *m/z* 42 channel, which is characteristic for $CH_2CO$, is also presented.

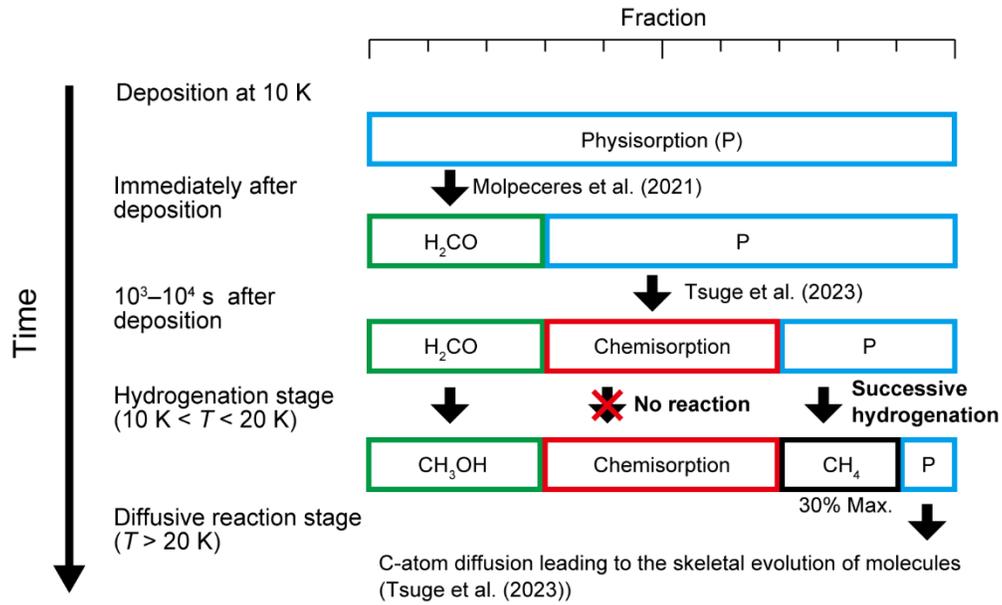

**Figure 4.** Schematic of the evolution of C atoms accreted onto icy grains at a low temperature (~10 K) under molecular cloud conditions. The fraction represents the probability of each reaction pathway and not the distribution of binding sites.